\documentclass[11pt,a4paper]{article}
\usepackage{jheppub}
\usepackage{epstopdf}

\usepackage{amssymb}
\usepackage{graphics,bm}
\usepackage{epsfig}
\usepackage{graphicx}
\usepackage{rotating}

\newcommand{\beq}{\begin{equation}}
\newcommand{\eeq}{\end{equation}}
\newcommand{\ba}{\begin{eqnarray}}
\newcommand{\ea}{\end{eqnarray}}

\def\square{\vcenter{\vbox{\hrule height.4pt
          \hbox{\vrule width.4pt height4pt
          \kern4pt\vrule width.3pt}\hrule height.4pt}}}

\title{Four results on $\phi^4$ oscillons in D+1 dimensions}

\author[a]{Erik Alexander Andersen,}
\date{today}
\author[a]{Anders Tranberg}
\emailAdd{anders.tranberg@nbi.dk}
\emailAdd{mjc615@alumni.ku.dk}
\affiliation[a]{
Niels Bohr International Academy, Niels Bohr Institute and Discovery Center,
Blegdamsvej 17, DK-2100 Copenhagen, Denmark}

\abstract{We present four results for oscillons in classical $\phi^4$ theory in D+1 space-time dimensions, based on numerical simulations. These include  the oscillon lifetime and the dependence on D; evidence for the uniqueness of the oscillon; evidence for the existence of oscillons beyond D=7; and a brief study of the spectrum of the radiation emitted from the oscillons before, during and after its ultimate demise.}

\begin{document}

\maketitle

\section{Introduction}
\label{sec:introduction}

Oscillons are non-topological, long-lived and quasi-periodic field configurations in certain non-linear field theories \cite{oscillons1,oscillons2,gleiser1,gleiser2,tuning1,tuning2,gleisergaussian,saffin,petja,petja2D,fodor1,gleiserdamping}. Much is known both analytically and numerically about these objects, but a complete understanding of their existence, creation and ultimate decay is still not settled. 

In the following, we will give at least partial answers to four questions one may want to ask about oscillons in general, but concentrating on a real, classical scalar $\phi^4$-theory. The action reads
\ba
\label{eq:action}
S=-\int d^{D+1}x \left[\frac{1}{2}\partial_\mu\phi\,\partial^\mu\phi-\frac{\mu^2}{2}\phi^2+\frac{\lambda}{4}\phi^4\right],
\ea
and it is known that the theory has oscillon solutions. The two degenerate minima are at $v=\pm\mu/\sqrt{\lambda}$ and the mass of excitations around these minima (the ``radiation'' frequency) is $m=\sqrt{2}\mu$. We will assume spherical symmetry in $D$ dimensions, and in this case the classical equation of motion reads:
\ba
\label{eq:eom}
\left[\partial_t^2-\partial_r^2-\frac{D-1}{r}\partial_r-\mu^2+\lambda\,\phi^2\right]\phi=0.
\ea
We will discretize and solve this equation on a spatial lattice in real-time. By assuming spherical symmetry, the lattice is in practice one-dimensional, with the physical dimensionality appearing only as the parameter $D$. This being the case, we will treat $D$ as a continuous parameter, i.e. allow for a non-integer number of dimensions.

In order that radiation emitted from the oscillon does not influence the later evolution of the oscillon, we will impose absorbing boundary conditions, generalizing the approach in \cite{petja} to D dimensions. This means that at the large-$r$ boundary, we replace (\ref{eq:eom}) by (see Appendix \ref{sec:A})
\ba
\left[\partial_t^2+\partial_r\partial_t+\frac{D-1}{2r}\partial_t+\mu^2\right]\phi=0.
\label{eq:boundary}
\ea
The alternatives are to have periodic boundaries and large lattices or some dissipative procedure far from the oscillon core \cite{gleiserdamping,saffin}.
At $r=0$, spherical symmetry imposes $\partial_r\phi=0$, and we therefore have
\ba
\left[\partial_t^2-D\partial_r^2-\mu^2+\lambda\,\phi^2\right]\phi=0.
\ea
We initialize the field with a Gaussian profile
\ba
\label{eq:initial}
\phi(r,0) = 1-C\exp\left(-\frac{r^2}{r_0^2}\right),
\ea
with two free parameters, the amplitude $C$ and the width $r_0$. The initial momentum is set to $\dot{\phi}(r,t=0)=0$, for all $r$.

We will take $\mu=1$ and $\lambda=1$, which amounts to a rescaling of field and space-time coordinates, so that $t\rightarrow \mu t$, $x\rightarrow \mu x$, $\phi\rightarrow \phi/v$. Then the minimum of the potential is at $\phi=\pm 1$, and lifetimes and sizes of the oscillons are in mass units throughout. The radiation frequency is then $m=\sqrt{2}$.

In the following, we will consider the evolution of the energy, as well as the instantaneous oscillation frequency of the center of the oscillon $\omega(t)$, taken to be 
\ba
\omega(t)=\frac{2\pi}{\delta t},
\ea
where $\delta t$ is the time between a crossing of $\phi(0,t)=0$ and the crossing a period (i.e. two crossings) later.

\begin{figure}
\begin{center}
\includegraphics[width=0.48\textwidth]{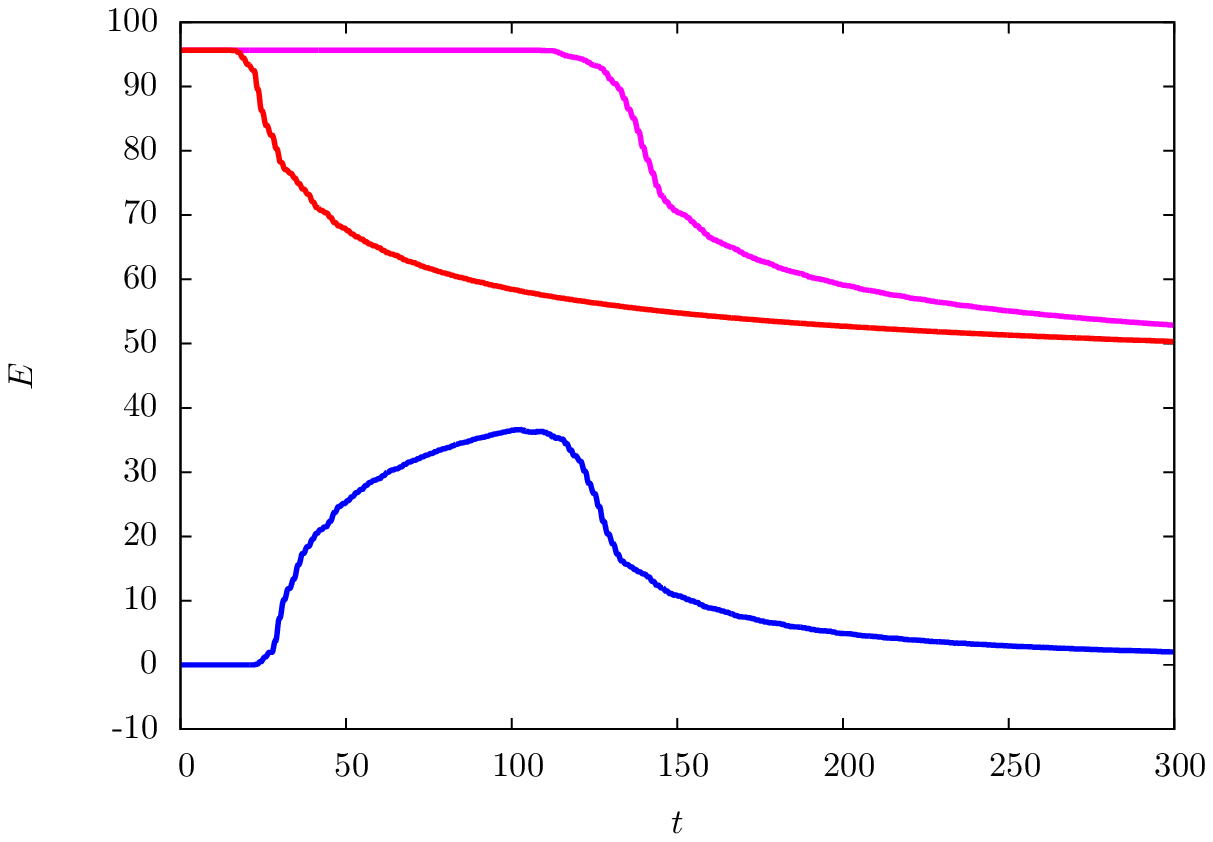}
\includegraphics[width=0.48\textwidth]{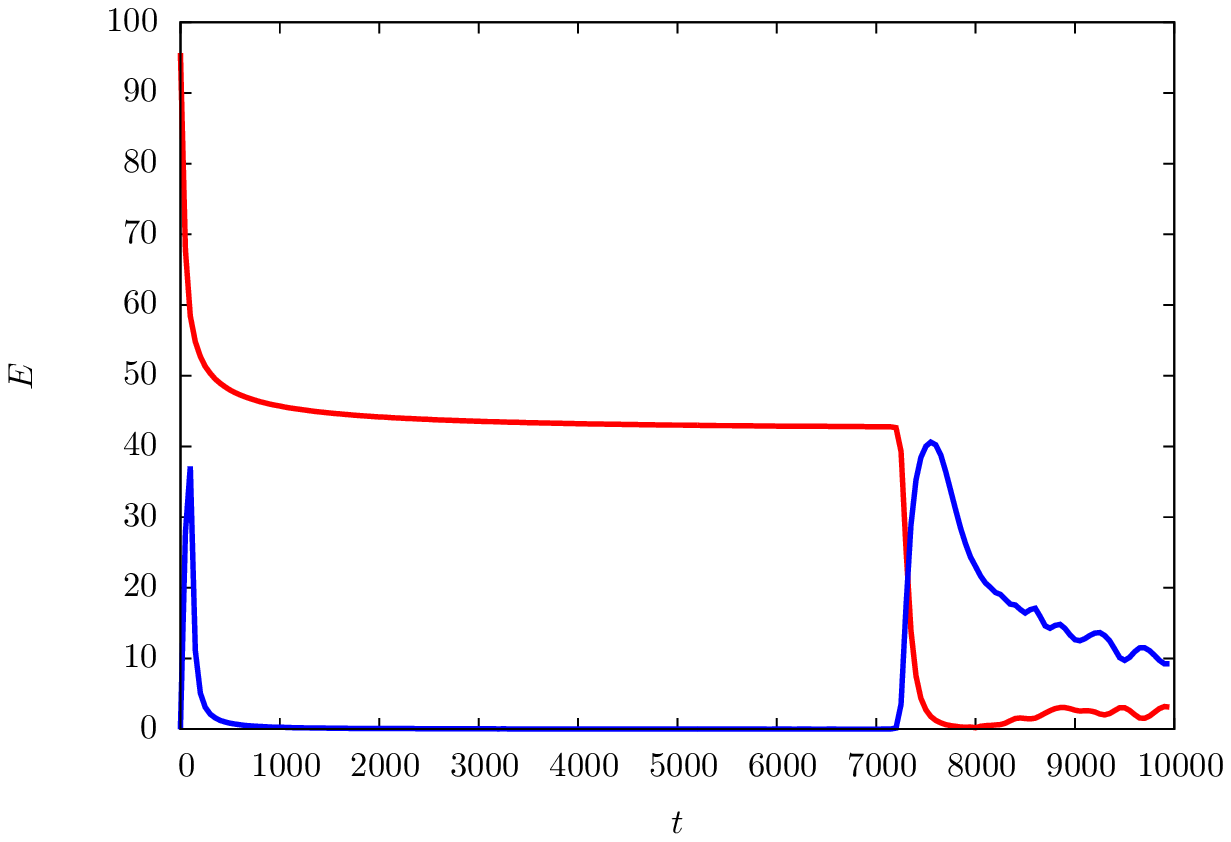}
\caption{The energy inside a box around the center (red), of the radiation outside this box (blue) and the total energy on the lattice (magenta), at short (left) and long (right) times. $C=-1$ and $r_0=3$.}
\label{fig:energyex}
\end{center}
\end{figure}

Fig.~\ref{fig:energyex} (left) is an example of the energy inside a box of radius $r=15$ around the oscillon (red), the radiation in the range $20<r<90$ (blue) and the sum over the whole lattice (magenta). At first the oscillon sheds energy into radiation (red curve drops, blue curve rises), but total energy is conserved. Then around $t=100$, the radiation reaches the absorbing boundary, and radiation and total energy drops, while the oscillon continues to pump energy into radiation. This is in fact only a transient stage until the Gaussian initial condition settles to a true oscillon at late times, with some almost constant energy. 

Fig.~\ref{fig:energyex} (right) is the later time evolution of the oscillon, we see the brief transient and the long oscillon stage, which ends suddenly at $t\simeq 7200$, where all the energy is dumped into radiation. We also note that during the oscillon stage, very little radiation is emitted, ad so very little radiation reaches the boundary.

\section{What is the lifetime of the oscillon in D dimensions?}
\label{sec:lifetime}

\begin{figure}
\begin{center}
\includegraphics[width=0.9\textwidth]{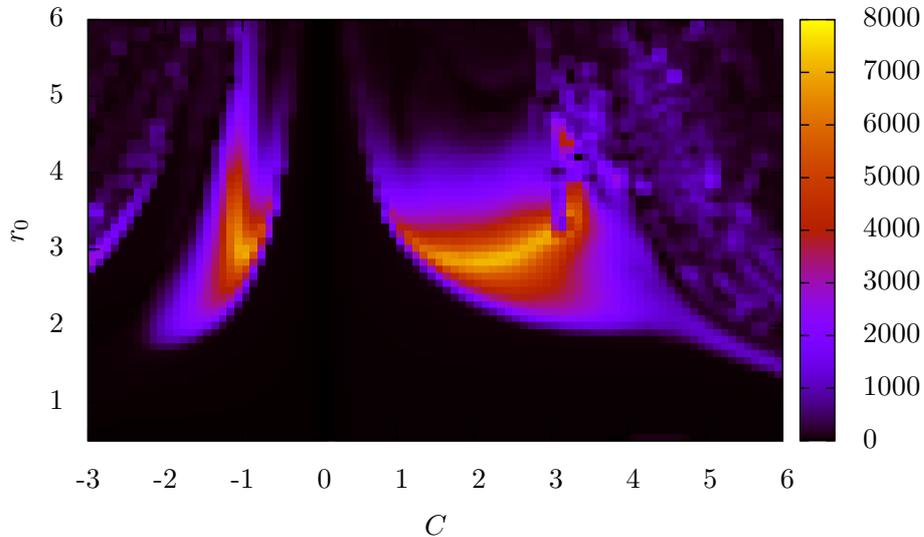}
\caption{The lifetime of oscillons from Gaussian initial conditions in $D=3$, in the $C$-$r_0$ plane.}
\label{fig:lifetime3}
\end{center}
\end{figure}

In \cite{saffin}, spherically symmetric simulations in D+1 dimensions were performed using dissipative dynamics away from the oscillon. Using absorbing boundary conditions allows much more efficient numerics, since we can do with smaller lattices. Also we find that this approach is better at getting rid of spurious waves bouncing back from the far boundary.

In Fig.~\ref{fig:lifetime3} we show the lifetime of the oscillon in $D=3$ as a function of the width and amplitude of the initial Gaussian. The lifetime is simply measured from the initial profile at $t=0$ until the oscillon collapses. We see two clear regions where the lifetime is large, up to $t=8000$, corresponding to starting the oscillon either up the potential (left side) or across to the other potential minimum and beyond (right side). The two parameter regions are of course connected by the oscillating behaviour of the oscillon, so that they should not really be seen as separate.

The black region is where the initial condition does not evolve into an oscillon, and it is clear that there is a lower limit in the amplitude, around the inflection point of the potential $C>1-\sqrt{1/3}\simeq 0.42$. Although more fuzzy, there also seems to be upper and lower limits to the region of large lifetime, say around $-2$ and $4$. The limits on the allowed width is strongly correlated with the amplitude, but in general $r_0>2$. Finally, there may be an upper limit to the width of roughly $r_0<5$, but this is less clear.

\begin{figure}
\begin{center}
\includegraphics[width=0.48\textwidth]{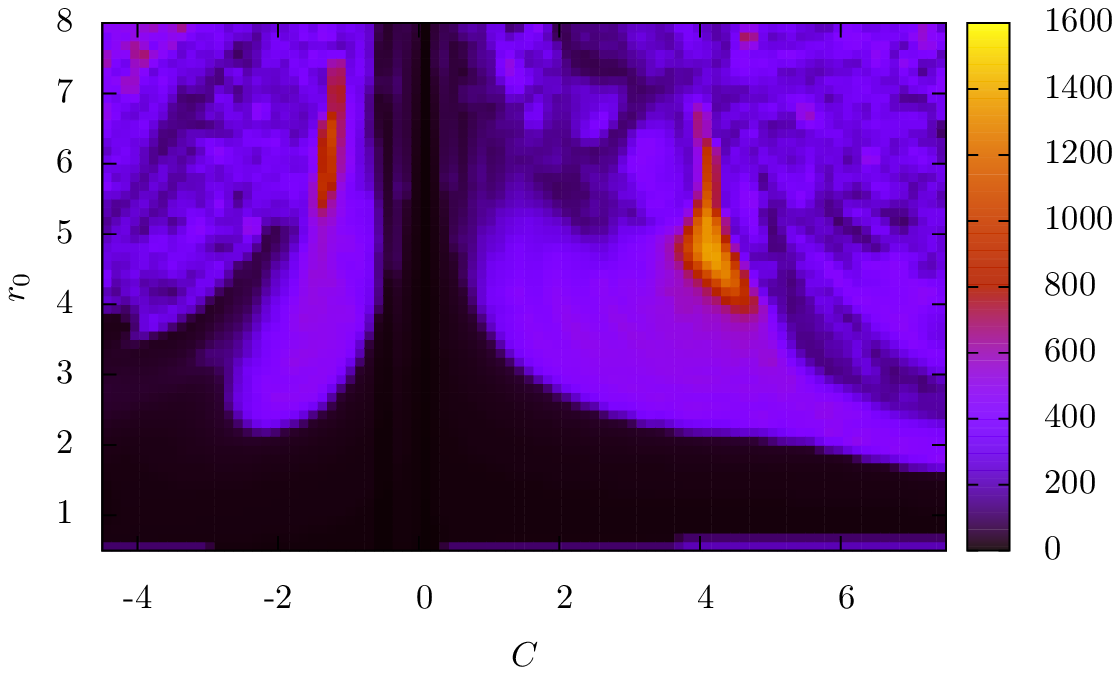}
\includegraphics[width=0.48\textwidth]{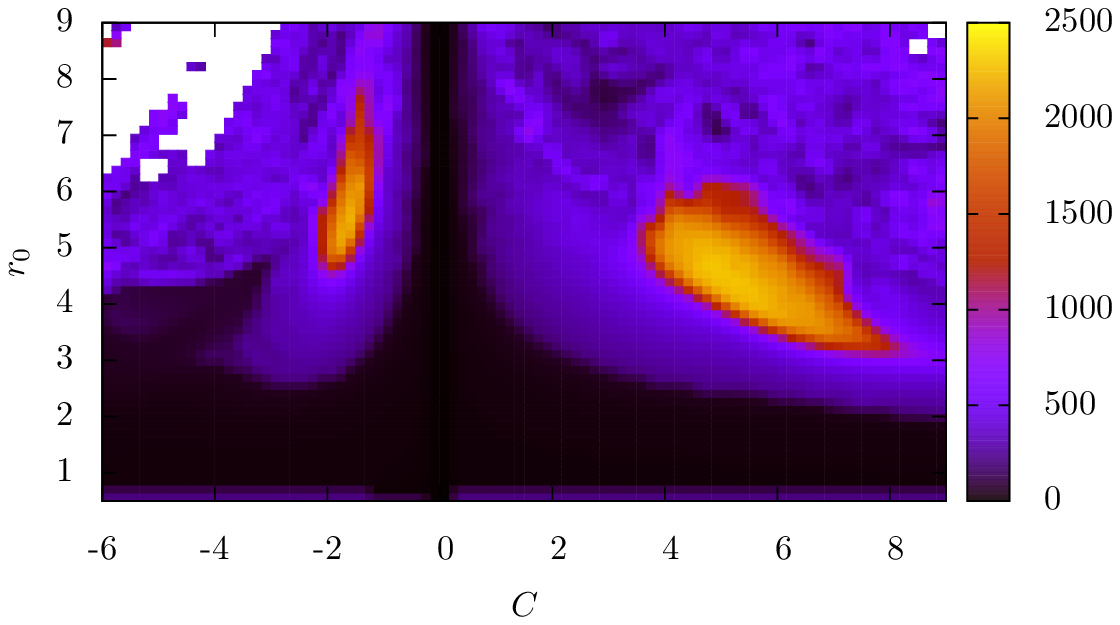}
\caption{The lifetime of oscillons from Gaussian initial conditions in $D=4$ (left) and $D=5$ (right), in the $C$-$r_0$ plane.}
\label{fig:lifetime45}
\end{center}
\end{figure}

In Fig.~\ref{fig:lifetime45} we show a similar result, but now for $D=4$ and $D=5$. We observe a similar picture of two main parameter regions, but whereas the lower limits to the basin of attraction are again quite sharp, the upper limits are more fuzzy. As in $D=3$, the lower limits are $r_0>2$, and the amplitude above the inflection point, larger for small width. But the maximum lifetime regions are at larger width and amplitude than for $D=3$.

\begin{figure}
\begin{center}
\includegraphics[width=0.9\textwidth]{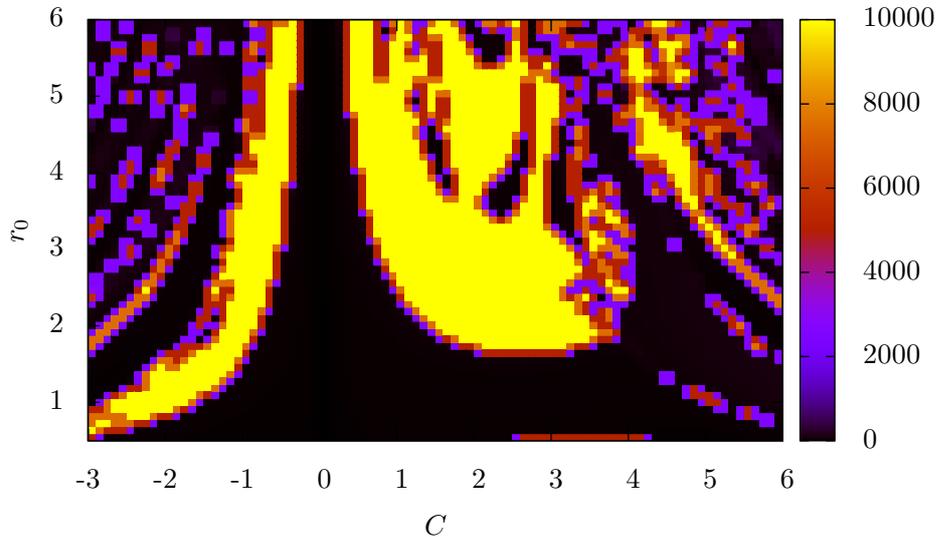}
\caption{The lifetime of oscillons from Gaussian initial conditions in $D=2$, in the $C$-$r_0$ plane.}
\label{fig:lifetime2}
\end{center}
\end{figure}
Perhaps surprisingly, the maximum lifetime in $D=5$ is larger than for $D=4$. In fact, because the lower limits shift, and the maximum lifetime regions move about, a given parameter set $C,r_0$ can fall out of the oscillon basin of attraction and into the collapse region (black area). In \cite{saffin}, the lifetime as a function of $D$ was reported to follow a power law until $D=5$, after which a second power law set in. In our language, the configuration considered there is $(C,r_0)=(2,3.3)$, which for $D=3$ is nicely in the maximum lifetime region, but which for $D=4,5$ falls off the basin of attraction. In \cite{saffin} the onset of the second power law was used to argue that oscillons do not exist for high $D$. More on this in section \ref{sec:largeD}.

In Fig.~\ref{fig:lifetime2} we show the $C$-$r_0$ plane for $D=2$. It is known that oscillons can live for millions of oscillations (see for instance \cite{petja2D}), and we simply cut off at $t=10000$, since we are interested in the structure of the phase space. We see that indeed two major regions establish themselve, with very long lifetimes, but now on the left-hand side, multiple ``ridges'' areopening up. Also for large enough amplitude, oscillons arise from very narrow initial Gaussians (small $r_0$). Again, from the point of view of a fixed parameter point (say $(C,r_0)=(2,3.3)$), the right-hand oscillon region has shifted down and left.

\begin{figure}
\begin{center}
\includegraphics[width=0.9\textwidth]{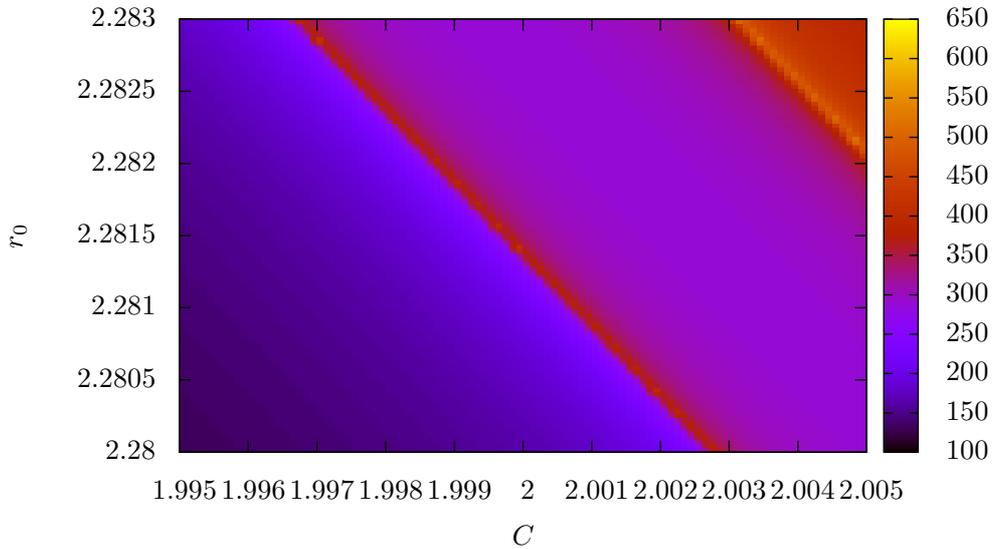}
\caption{The lifetime in $D=3$ in a narrow, but highly resolved region the $C$-$r_0$ plane, showing a fine band structure.}
\label{fig:tunelifetime}
\end{center}
\end{figure}

As was beautfifully demonstrated in \cite{tuning1,tuning2}, the lifetime as a function of width and amplitude in fact has a spiky structure, where by carefully tuning the parameters, one can get perhaps arbitrarily large lifetimes. In Fig.~\ref{fig:tunelifetime}, we show with much higher resolution the region near the point $r_0=2.28$, $C=2$ in $D=3$, which is one instance of the spiky structure reported in \cite{tuning1,tuning2}. In fact, the substructure is a set of very narrow bands, and indeed factors of 3 or more in lifetime can be achieved by careful tuning. As we see, the tuning has to be extremely fine (here less than one per mille) to see the additional structure. The broader behaviour of the lifetime is of course represented by the coarser sampling of figures~\ref{fig:lifetime3}-\ref{fig:lifetime2}, and a population of oscillons created during a phase transition are likely to have these lifetimes \cite{amin1,amin2,gleiser3,broadhead1}, but the fine structure is a signal of the non-trivial nature of the dynamics, and deserves closer scrutiny in its own right.

\section{Is there more than one oscillon?}
\label{sec:uniqueness}

\begin{figure}
\begin{center}
\includegraphics[width=0.48\textwidth]{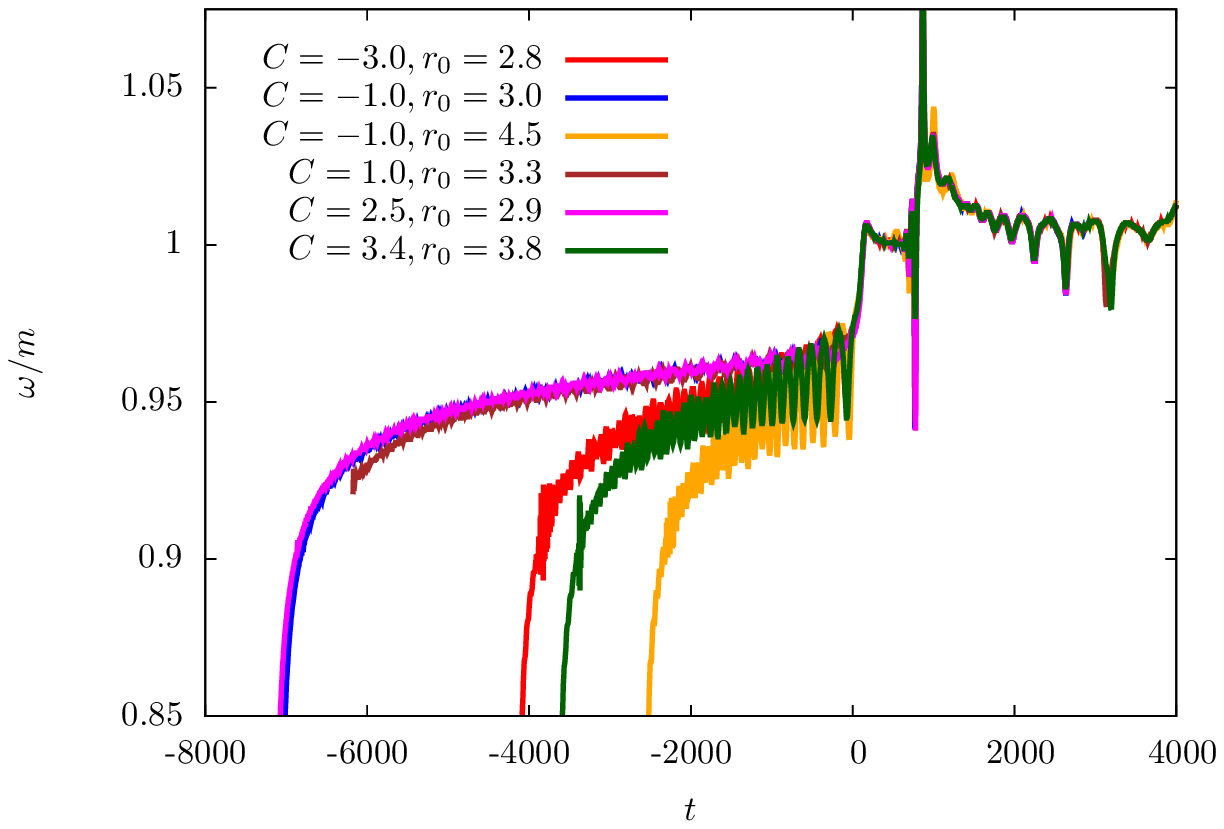}
\includegraphics[width=0.48\textwidth]{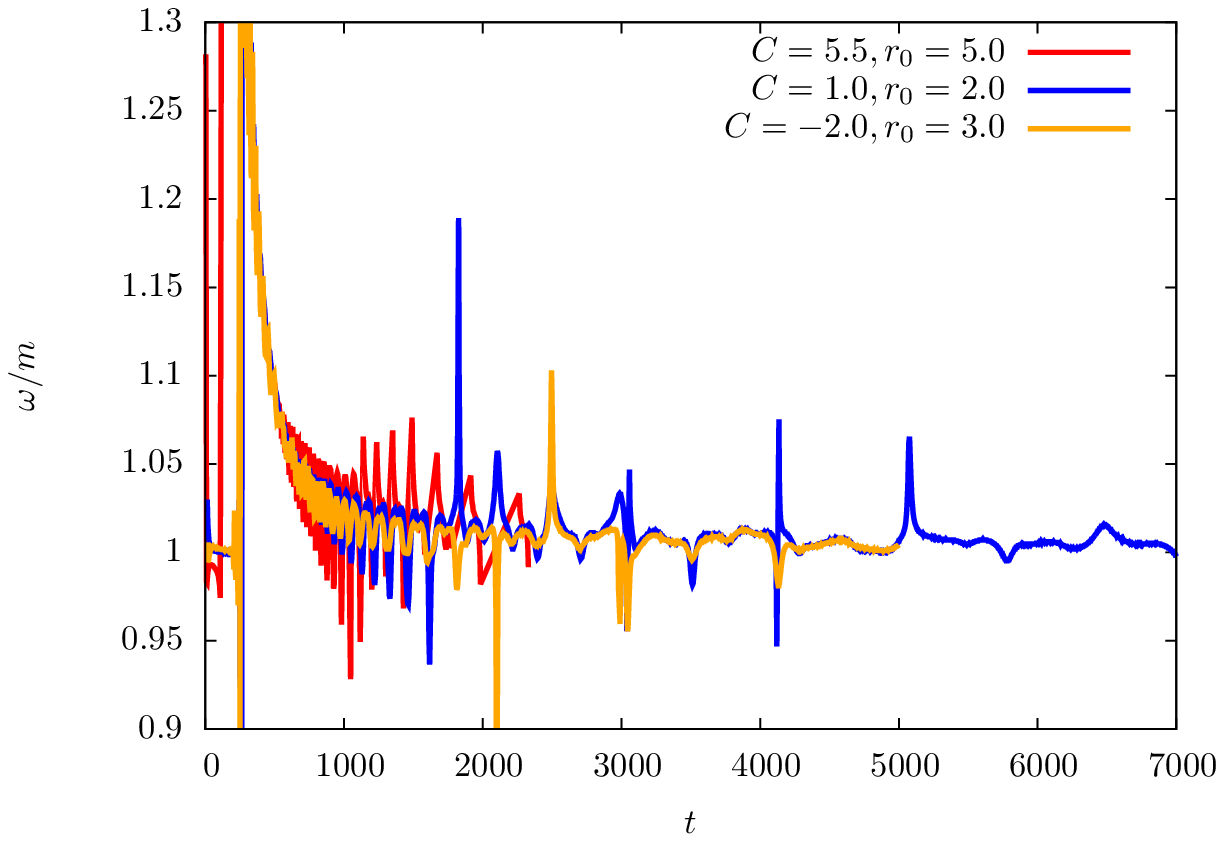}
\caption{A number of trajectories of $\omega(t)$ starting from different Gaussian initial conditions in $D=3$. Left: Initial conditions leading to oscillons, Right: Initial conditions in the black region, where there is immediate collapse to radiation.}
\label{fig:attractor}
\end{center}
\end{figure}

The current understanding of oscillons is that they are attractor solutions to the equations of motion, in the sense that within a basin of attraction, all initial localized field configurations approach some oscillon solution. we have seen in the previous section how these regions manifest themselves in terms of the lifetime.

The approach to a true oscillon happens by an initial shedding of energy as radiation and the changing of shape to the oscillon profile. The oscillon performs a non-linear oscillation with a slowly changing frequency $\omega(t)$ slightly below the mass $m=\sqrt{2}\mu$, which is why it cannot easily decay into free excitations of the field. 

Oscillons slowly evolve by emitting a little radiation and changing their frequency to a critical value $\omega_c$, where they collapses into only radiation at $\omega=m$ (see for instance \cite{saffin}). It is therefore natural to think of an oscillon as a whole trajectory of field configurations ending at $\omega_c$, and different initial configurations in the basin of attraction will asymptote to different points along this trajectory.

Now the question is: Is there more than one oscillon trajectory in $\phi^4$ theory in $D+1$ dimension? In other words, can the field space (of localized energy configurations) be split up into only two basins of attraction, one leading to {\it the} oscillon (trajectory) and one leading to immediate collapse into radiation?

The first hint that the oscillon is unique is that the basins of attraction are dominated by two large regions, corresponding to the extremes of the oscillation (Figs.~\ref{fig:lifetime3}-\ref{fig:lifetime2}). Fig.~\ref{fig:attractor} (left) shows the trajectories of $\omega(t)$ of the oscillon center $\phi(0,t)$, for a number of Gaussian initial conditions, shifted so that the final collapse is simultaneous. The initial conditions are picked from different mutually disconnected regions in $D=3$. We see how they asymptote either to the same, unique trajectory, for the short-lived with an additional oscillation due to the oscillon radius ``wobbling''. In particular the evolution after collapse is remarkably similar. 

In Fig.~\ref{fig:attractor} (right) we picked a number of points from the black collapse region, and we see a completely different behaviour. $\omega(t)>m$ throughout, and it is not even clear that there is a well-defined oscillation. At late times the $\omega$ again asymptotes to $m$. Obviously, we cannot claim that gaussian initial conditions sample all configuration space, but the results suggest that the oscillon trajectory is unique, and that in this sense, there is only one oscillon. We checked that a similar picture emerges in $D=2,4$ and $5$. 

\section{Can there be oscillons in a large number of dimensions?}
\label{sec:largeD}

\begin{figure}
\begin{center}
\includegraphics[width=0.75\textwidth]{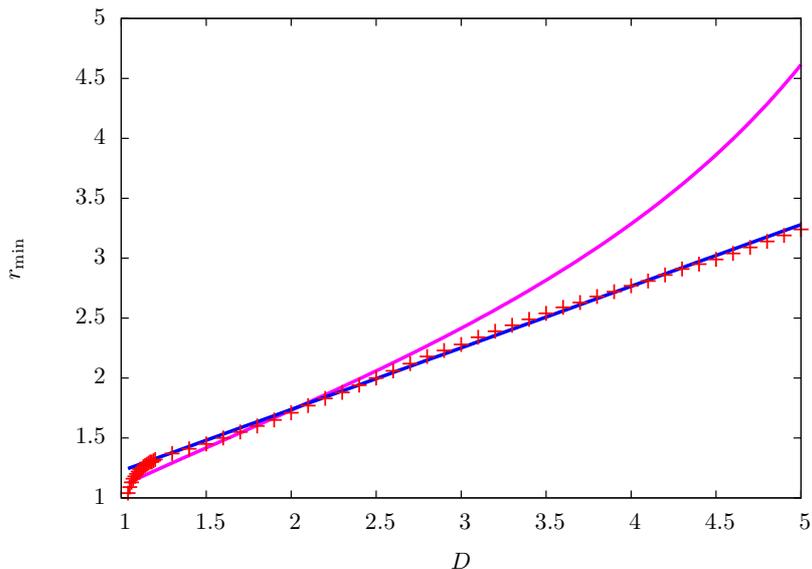}
\caption{The minimum width $r_{\rm min}$ allowing for an oscillon at fixed amplitude $C=2$. Superposed, the Gaussian estimate (magenta) and a linear fit $r_{\rm min}=(0.512\pm0.004)D+(0.71\pm0.01)$ (blue).}
\label{fig:largeD}
\end{center}
\end{figure}

In \cite{gleisergaussian}, it is suggested that oscillons cannot exist for a large number of spatial dimensions. This is based on analytical calculations for gaussian profiles, which show that there is a minimum Gaussian width $\sigma_c(D)$ which allows for oscillons,
\ba
r_{\rm min}=\sqrt{\frac{D}{3(2^{3/2}/3)^D-2}},
\ea
and since this diverges around $D\simeq 6.88$, no oscillons can exist for $D=7$ and beyond.

As shown in Fig.~\ref{fig:largeD}, at least the value of this limit must be put down to restricting not only the initial condition, but the whole evolution to a Gaussian anzats. We show the minimum Gaussian width leading to an oscillon at $C=2$, i.e. the smallest width along a particular slice in $C$-$r_0$ space, as a function of $D$. We see that although this quantity increases with $D$, there is no sign of any divergence near $D=7$. We have superposed the Gaussian analytic result of \cite{gleisergaussian} and a linear fit. It is however true that because the lifetime becomes smaller and smaller, ultimately it is hard to decide whether the oscillon configuration has the time to establish itself, before it collapses. But this is an artifact of the Gaussian initial condition, rather than the absence of a quasi-breather solution (see for instance \cite{saffin}).

\begin{figure}
\begin{center}
\includegraphics[width=0.48\textwidth]{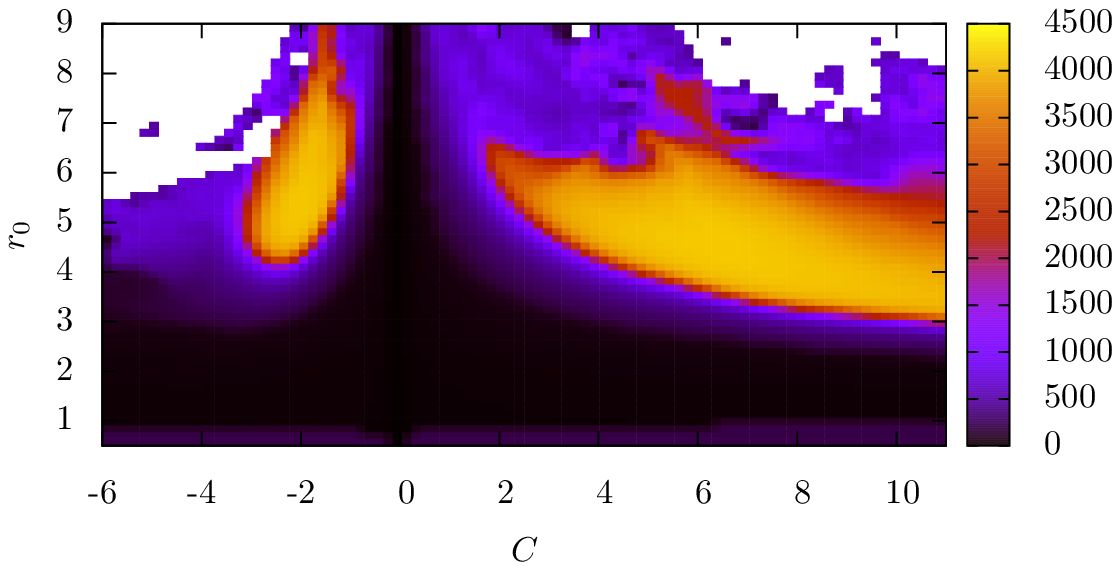}
\includegraphics[width=0.48\textwidth]{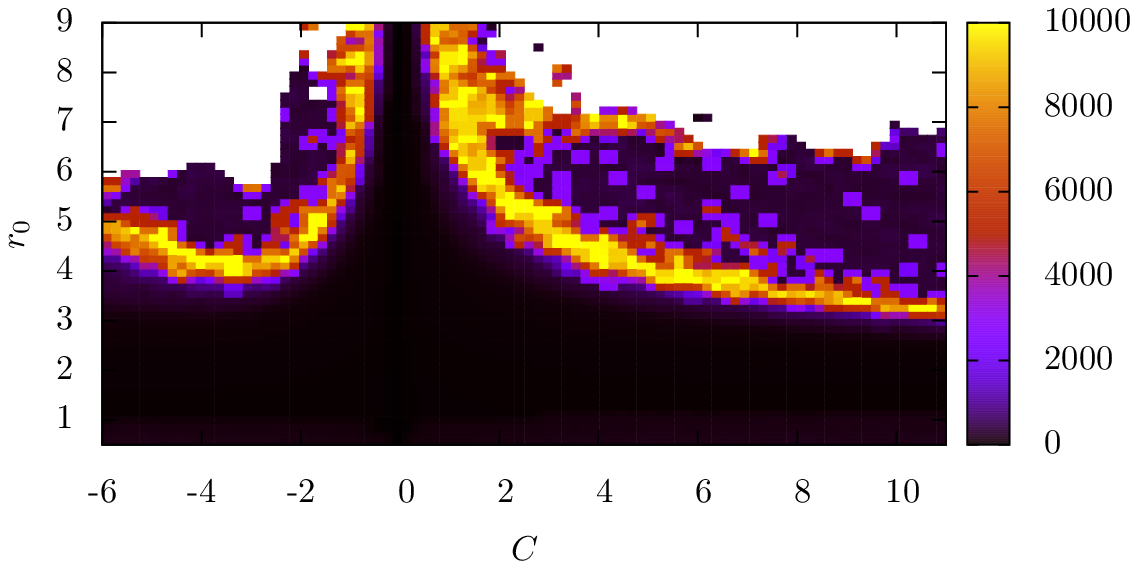}
\caption{The lifetime of oscillons from Gaussian initial conditions in $D=6$ (left) and $D=7$ (right), in the $C$-$r_0$ plane.}
\label{fig:lifetime67}
\end{center}
\end{figure}

Fig.~\ref{fig:lifetime67} shows the lifetime in the $C$-$r_0$ plane for $D=6$ (left) and $D=7$ (right). We see that there are certainly oscillons of significant lifetime. We should note that the left-out regions in the upper left and right-hand corners suffer from numerical instabilities which we were unable to cure. This becoms worse at higher $D$, and prevents us from simulating beyond $D=7$. But the two bulk regions and their lower boundary to the no-oscillon basin are physical. In $D=7$, these are rather narrow, and it is possible that there is an upper limit to $D$, or that only highly tuned bands remain.

\section{What happens when an oscillon dies?}
\label{sec:death}

At the end of its life, the oscillon suddenly collapses and all its energy is emitted into radiation, free field fluctuations peaked at $\omega=m$. As was also remarked in \cite{petja}, the non-linear oscillation of the oscillon can be resolved by Fourier transform into a discrete set of frequency peaks at integer multiples of the basic frequency. And so although the basic frequency is below the mass threshold, the higher modes are not. These modes are in fact responsible for the slow emission of radiation during the oscillon stage, as was shown in \cite{saffin,petja} (see also \cite{fodor1}). But because the secondary peaks are exponentially suppressed, very little radiation is emitted. 

\begin{figure}
\begin{center}
\includegraphics[width=0.48\textwidth]{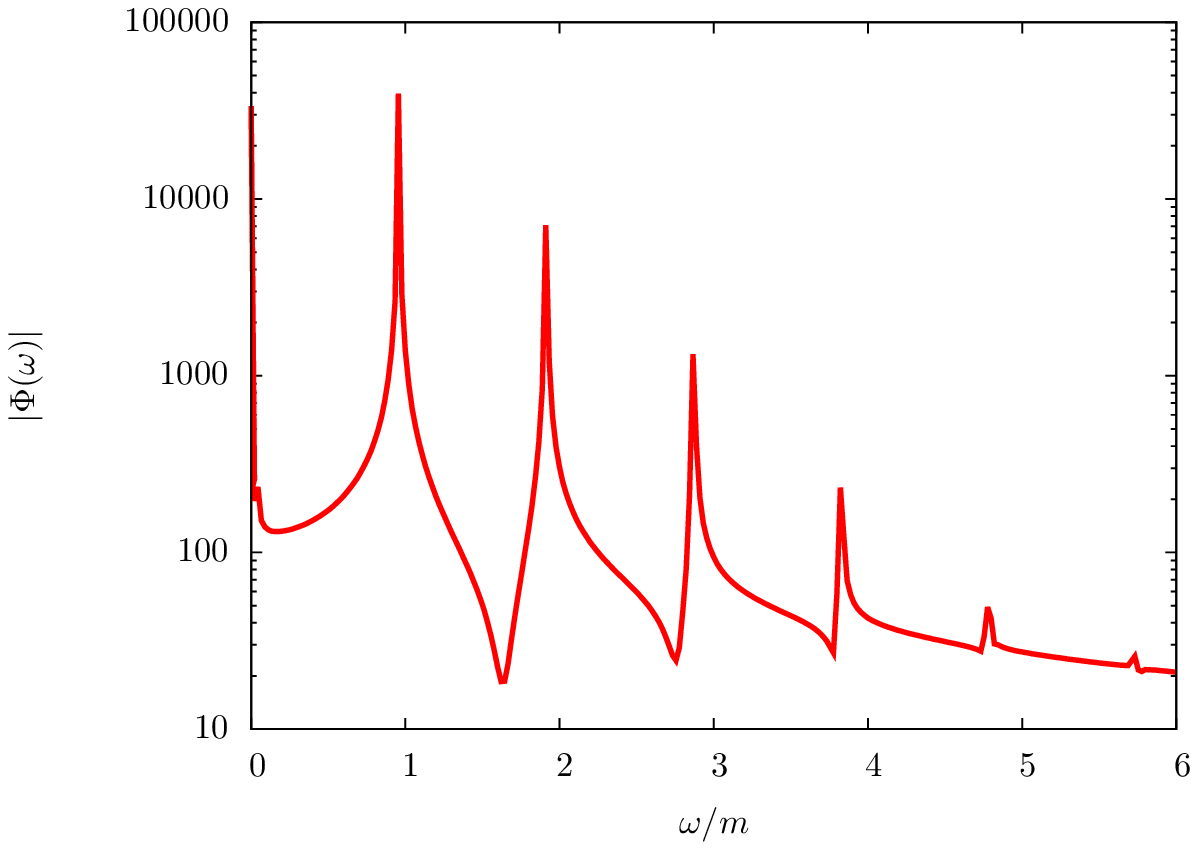}
\includegraphics[width=0.48\textwidth]{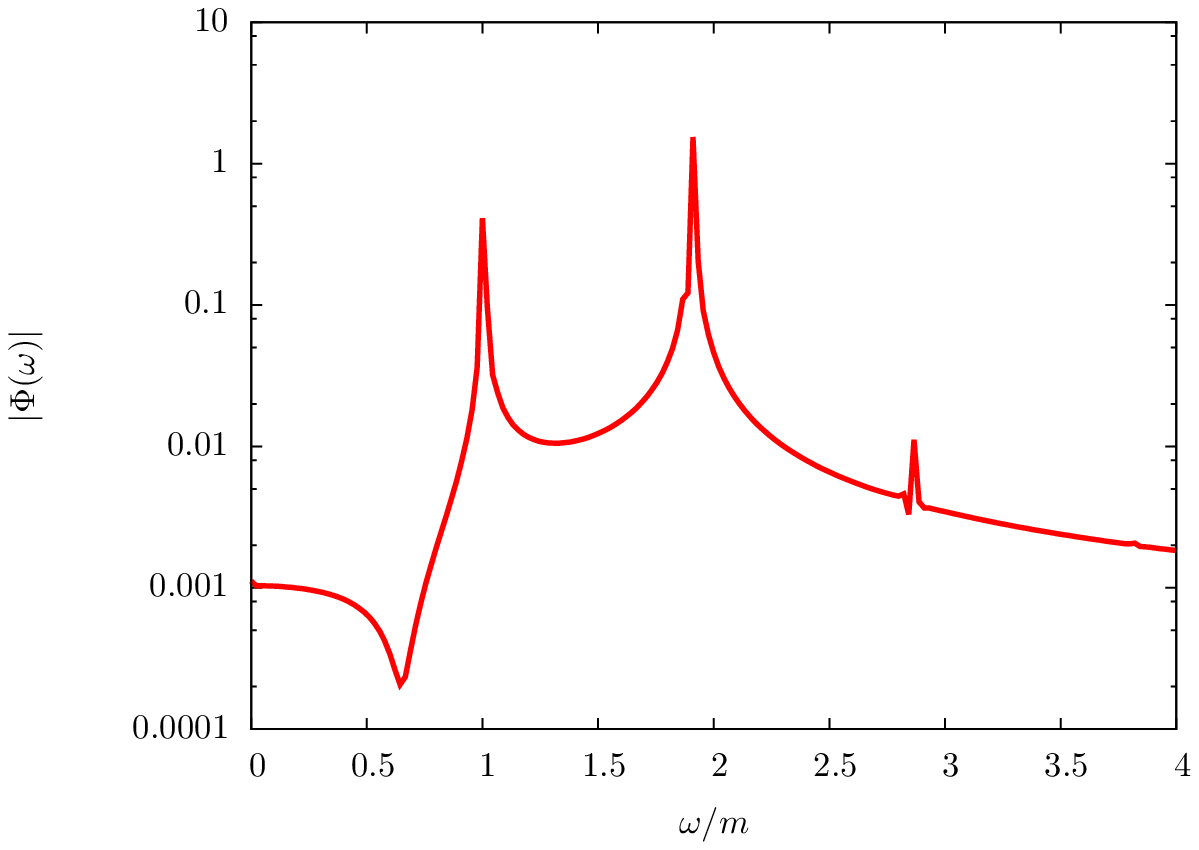}
\caption{Left: The Fourier transfrom of the oscillation in the center of the oscillon $r=0$ around time $4000$. Right: The same thing far from the center, $r=80$.}
\label{fig:radbefore}
\end{center}
\end{figure}

In Fig.~\ref{fig:radbefore} (left) we first show the Fourier transform in time of the center of the oscillon, $r=0$, in a time interval around $t=4000$. The initial condition is the same as for Fig.~\ref{fig:energyex}. We see that the dominant frequency is indeed slightly below the radiation frequency, $\omega_0/m=0.955$, and that there are exponentially suppressed peaks at integer multiples of this. Similarly, in Fig.~\ref{fig:radbefore} (right), we show the frequencies present far into the radiation domain, $r=80$. Now the largest peak belongs to the $2\times\omega_0$-mode, whereas the basic frequency is present only through the part of its peak that reaches beyond $\omega/m=1$. And so the width of the peaks is important, corresponding to the oscillation not being exactly periodic.

\begin{figure}
\begin{center}
\includegraphics[width=0.48\textwidth]{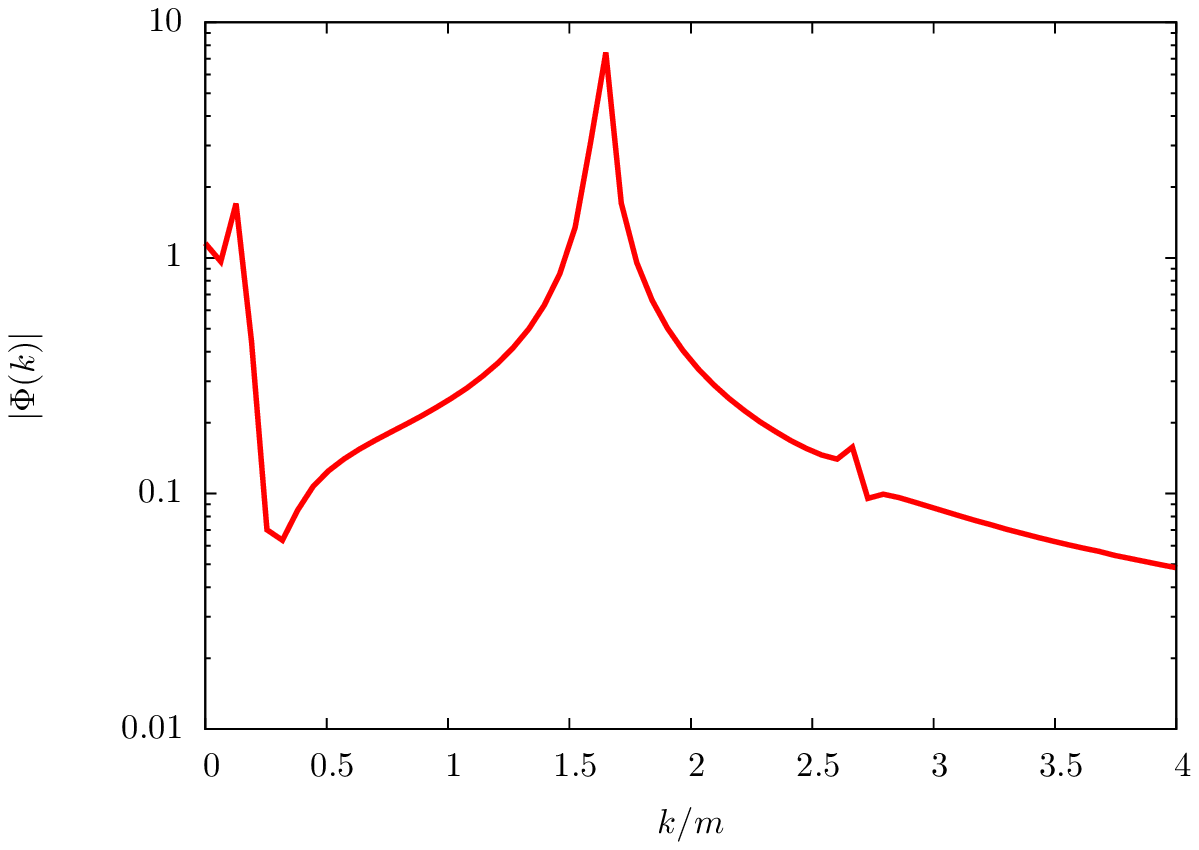}
\includegraphics[width=0.48\textwidth]{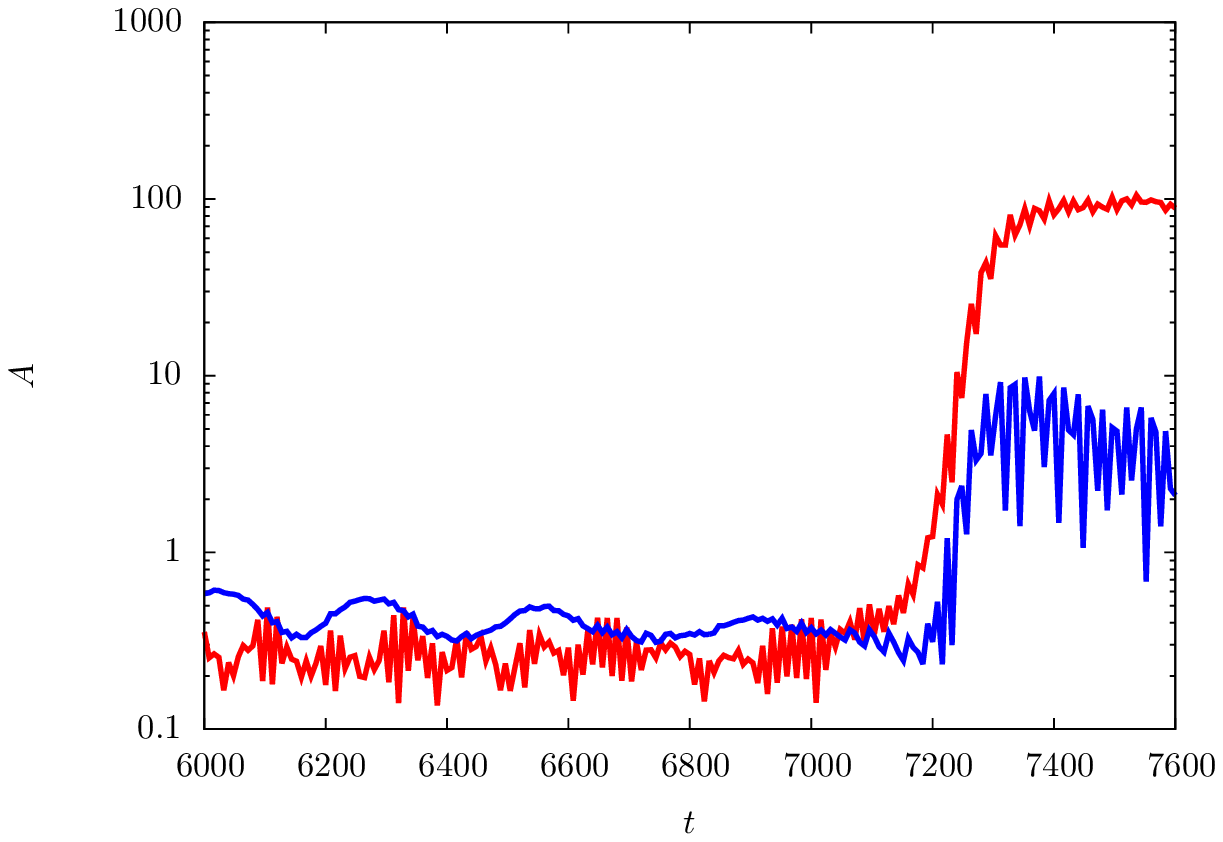}
\caption{Left: The spatial Fourier transform in the radiation region at time $t=4100$. Right: The integrated power in the $2\omega_0$ peak (blue) and the radiation peak at $\omega=m$ (red).}
\label{fig:raddecay}
\end{center}
\end{figure}

In Fig.~\ref{fig:raddecay} (left), we show the spatial ($r$-space) Fourier transform in the radiation region $20<r<90$. The peaks in $k/m$ correspond exactly to the $\omega$ in Fig.~\ref{fig:raddecay} through $\omega^2=m^2+k^2$, and we again observe the dominant $2\omega_0$-peak and the smaller $\omega_0$-tail near $k=0$. We can now track the amount of radiation in each of these peaks (near $2\omega_0$ and $m$) by integrating up the area under each peak and displaying their evolution in time. This is shown in Fig.~\ref{fig:raddecay} (right). We see that deep in the oscillon regime, the $2\omega_0$ peak dominates the radiation. But as early as $t=7150$, the radiation peak takes over. This precedes the collapse of the oscillon by $\simeq 100$ in time units, as can be seen from the energy in Fig.~\ref{fig:energyex}.

This means that the death of the oscillon is not an instantaneous process, but is preceded by a change in the radiation pattern. Oscillons decay because the width of the primary frequency peak near $\omega_0$ eventually opens up enough that energy can be emitted through it. 
 
\section{Conclusion}
\label{sec:conclusion}

We have considered four separate, but related, features of the oscillon in classical $\phi^4$-theory in $D$ dimensions. We have seen that 1) the oscillon is likely to be unique in the sense of there being only a single oscillon trajectory in $\omega$-space, or one non-trivial basin of attraction; 2) we have computed the lifetime for Gaussian initial conditions in $D=2-7$ and shows a two-region basin-of-attraction structure, the boundaries and maxima of which however shift as a function of $D$. Highly tunable lifetime ``spikes'' are in fact bands, in extremely small regions of parameter space; 3) there is no sign that oscillons cannot exist in $D>7$, or that their lifetime should be short. In fact the maximum lifetime as a function of $D$ has a minimum around $D=4$; 4) the eventual collapse of the oscillon is preceded and triggered by the primary oscillation frequency becoming efficient at radiating away energy. And therefore not only the position of the peak in the spectral function but also the width is of importance for the stability of oscillons. 

\acknowledgments
A. T. is supported by the Carlsberg Foundation and the Villum Kann Rasmussen Foundation.

\appendix

\section{Absorbing boundary conditions in D dimensions}
\label{sec:A}

We consider the field far from the center of the oscillon, in terms of the displacement away from the minimum, $\varphi=\phi-1$. The linearized equation of motion reads
\ba
\left[\partial_t^2-\partial_r^2-\frac{D-1}{r}\partial_r+m^2\right]\varphi=0, \qquad m^2=2\mu^2.
\label{eq:linearized}
\ea
Separating variables using $\varphi(r,t)=R(r)\exp(-i\omega t)$ allows us to write (\ref{eq:linearized})
\ba
\left[r^2\partial_r^2+(D-1)r\partial_r +(\omega^2-m^2)r^2\right]=0.
\ea
Using further $R(r)=r^{(2-D)/2}S(r)$ and $x=kr=\sqrt{\omega^2-m^2}$, we get the Bessel equation
\ba
x^2S''+S'\left(x^2-\frac{1}{4}(D-2)^2\right)S=0.
\ea
In the limit $r\gg1$, $x\gg 1$ this has the solution $x^{-1/2}\exp(ix)$, so that finally
\ba
\varphi(r,t)=r^{(1-D)/2}\exp\left[i(kr-\omega t)\right].
\label{eq:radiation}
\ea
We now note that the outgoing wave $\omega>0$ obeys
\ba
\left(\partial_r-ik+\frac{D-1}{2r}\right)\varphi=0.
\ea
Multiplying this by $\omega$, expanding $k$ to first order in $m^2/\omega^2$ and making the replacement $-i\omega\rightarrow \partial_t$, we arrive at (\ref{eq:boundary}).



\end{document}